\documentclass[multphys,vecphys]{svmult}

\usepackage{makeidx}         % allows index generation
\usepackage{graphicx}        % standard LaTeX graphics tool
\usepackage{multicol}        % used for the two-column index
\usepackage[bottom]{footmisc}% places footnotes at page bottom
\makeindex             % used for the subject index
\begin{document}

\title*{Gamma-ray probe of the QSO's obscured evolution}
\author{Anatoly Iyudin\inst{1}\and
Jochen Greiner\inst{2} \and Guido Di Cocco\inst{3} \and Stefan Larsson\inst{4}}
\institute{SINP, Moscow State University,
  Moscow, Russia
\texttt{aiyudin@srd.sinp.msu.ru}
\and MPI f\"ur extraterrestrische Physik, Garching, Germany \texttt{jcg@mpe.mpg.de}
\and IASF/INAF-Bologna, Bologna, Italy
\texttt{dicocco@bo.iasf.cnr.it}
\and AlbaNova Center, Stockholm University, Sweden
\texttt{stefan@astro.su.se}
}
\maketitle

\begin{abstract}
The obscured phase of QSOs, as well as their accretion history, can be best followed by observing QSOs bright in $>$10 MeV gamma-rays. By analysing the resonant absorption troughs in spectral energy distribution of flaring QSOs one can measure the (baryonic) absorbing column and baryonic content of the QSO host galaxy, while the flare strength will give information on the accretion rate of the QSO powering supermassive black hole. By measuring the baryonic absorbing column for QSOs at different redshifts one can follow the early obscured evolution of AGN at redshifts up to z$\sim$6.

\end{abstract}

\section{Obscured AGNs}
\label{iyudin:sec:1}
% Always give a unique label
% and use \ref{<label>} for cross-references
% and \cite{<label>} for bibliographic references
% use \sectionmark{}
% to alter or adjust the section heading in the running head
The idea that 
substantial absorption in AGN could be a definite characteristic of the early phases 
of QSO evolution [2] was invoked to explain the submillimeter observations of X-ray absorbed AGN at z$\sim$1 -- 3, that have shown strong emission at 850 $\mu$m [10]. 
This is a signature of copious star formation. In AGN evolution model, the main obscured growth phase of 
the QSO coincides with formation of the host galaxy spheroid, the completion of which indicates the beginning of the luminous, unobscured phase of the QSO's evolution [12]. The verification of this model at redshifts up to z$\sim$6 appears possible via application of the $\gamma$-ray resonant absorption method.

\subsection{Gamma-ray absorption}
\label{iyudin:sec:2}
The recently introduced absorption method based on the detection of the resonant absorption troughs in the gamma regime [5] has all the qualities to become a 
practical tool to measure baryonic absorption columns along the line-of-sight from the point source towards the observer. The pencil-like $\gamma$-ray beam from the QSO (or GRB) probes all absorbers along the line-of-sight providing their appropriate columns and redshifts, including the QSO's (GRB) host galaxy, as well as the matter in the Milky Way halo [6].
This $\gamma$-ray absorption method relies on the resonance-like photoabsorption by atomic nuclei. 
The photoabsorption on nuclei have three peaks in the cross section, at energies of $\sim$7 MeV, in the region of the ``pygmy'' dipole resonance (PDR), at 20-30 MeV (giant dipole resonance (GDR)), and at $\sim$325 MeV (${\Delta}$-resonance), see references to the relevant processes in [5].  
\par
Best studied of the above three processes are the GDR, and $\Delta$-isobar resonances, with the typical absorption
cross sections per nucleon of ${\sigma}_{GDR}$$\sim$1.1 mb, and
${\sigma}_{\Delta}$$\sim$0.5 mb.
The $\Delta$-isobar resonance has the energy of the absorption peak in the cross section, and the cross section peak value per nucleon, which is the same for all nuclei.
The GDR cross section peak energy varies with different nuclei, but for the solar mixture of elements it is primarily defined by the properties of GDR on $^4$He [5]. Either of GDR or $\Delta$-isobar resonances, can be used to derive the absorbing column value. 

In the $\gamma$-ray absorption experiment we have a rather simple geometry where a pencil-like beam of $\gamma$-ray photons have to pass through one or many absorbers on the way from a point source (QSO) towards an observer.
Therefore the differential photon flux observed at the Earth can be written as a  function of the photon energy and of redshift:
\begin{equation}
\frac{dN}{dE}= ( \frac{dN}{dE} )_{unabsorbed}\cdot e^{-\tau(E,z)} \hspace{2mm}.
\end{equation}
The dependence on $E$ and $z$ are quite complex, to simplify it we assume that we are dealing with two absorbers, one in the QSO host galaxy, and the second absorber in the Milky Way. To derive the absorbing column(s) one has to fit the QSO's spectral energy distribution, including troughs.

Absorption columns of a given quasar derived from the UV or optical observations are always
$\sim$100 times smaller than the X-ray derived columns (see [1] and Fig. 1 (left)). Similarly, from the same kind of relation, absorbing columns derived in the $\gamma$-ray regime are larger than those derived via X-ray measurements (Fig. 1(right)).

The absorption columns of $\gamma$-ray bright quasars found with the $\gamma$-ray absorption method [5, 6] are of the order of 10$^{26}$ cm$^{-2}$. With the superior sensitivity of {\it GLAST} absorbing columns as low as $\sim$10$^{25}$ cm$^{-2}$ will become measurable and thus close the gap (Fig. 1) between the X-ray and gamma-ray column distributions.
The absorption columns and redshifts of $\gamma$-ray bright quasars detected by {\it GLAST} will also enable us to follow the evolution of the halo mass of QSO hosting galaxies up to redshifts of z$\approx$6.
\begin{figure}
% For example, with the option graphics us
\centering
\includegraphics[width=11.4cm]{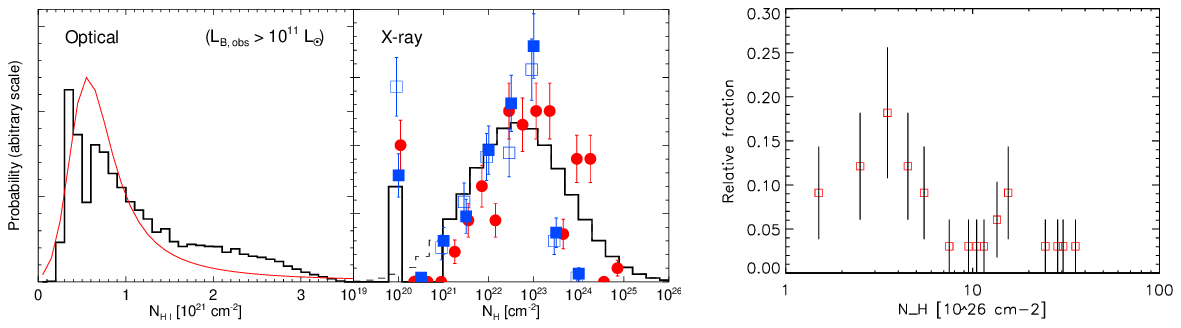}
%\picplace{5cm}{2cm} % Give the correct figure height and width in cm
%
\caption{Left: Typical distributions of the absorbing column densities derived towards luminous quasars in optical samples (left), and from the hard X-ray sample (right). Figure is adopted from Hopkins et al. (2005). X-ray data shown are from Treister et al. (2004) (blue squares) and from Mainieri et al. (2005) (red circles). Right: Column densities of absorbers in QSO hosting galaxies and in the Milky Way halo derived by the gamma-ray absorption method from SEDs of QSOs and EUIDs detected by EGRET.}
\label{iuydin:fig:1}       % Give a unique label
\end{figure}
\section{Conclusions}
\label{iyudin:sec:3}

Assuming that gravitational coupling of the baryonic and the Dark Matter holds [7], it is possible to use the measured absorption columns to probe the evolution of the baryonic matter content in the halo of QSO host galaxies.
Different constituents of the Dark Matter halo can be traced via the gamma-ray absorption method:
(1)- ordinary baryonic matter, like hydrogen, helium etc..., that constitutes the cold globules of Pfenniger (2004) [11]; (2)- dense color superconducting clumps [9]; (3)- dark baryons, relatives of X-particle [14]; or (4)- mixture of heavy and light eigenstates.

The $\gamma$-ray absorption method can provide an info on the evolution of ${{\Omega}_b}$ starting from the beginning of the reionization epoch. The sensitivity and an energy range of {\it GLAST} will allow to follow evolution of the baryonic halo of QSO host galaxies up to z$\approx$6. To probe the proto-galactic haloes at z$\approx$15 one will need the $\sim$1 MeV to 500 MeV range telescope with the sensitivity comparable or exceeding that of the {\it GLAST}. 
This energy range is suitable to cover both resonance-like photoabsorption troughs at $\sim$25/(1+z) MeV (GDR) and 325/(1+z) MeV ($\Delta$-isobar), and to have a few energy bins outside of the absorption troughs to constrain the continuum shape. 


%

%
% BibTeX users please use
% \bibliographystyle{}
% \bibliography{}
%
% Non-BibTeX users please follow the syntax
% the syntax of "referenc.tex" for your own citations
%%%%%%%%%%%%%%%%%%%%%%%% referenc.tex %%%%%%%%%%%%%%%%%%%%%%%%%%%%%%
% sample references
% "physics"
%
% Use this file as a template for your own input.
%
%%%%%%%%%%%%%%%%%%%%%%%% Springer-Verlag %%%%%%%%%%%%%%%%%%%%%%%%%%

%
% BibTeX users please use
% \bibliographystyle{}
% \bibliography{}

%%%%%%%%%%%%%%%%%%%%%%%%%%%%%%%%%%%%%%%%%%%%%%%%%%%%%%%%%%%%%%%%%%%%%%  }

%%%%%%%%%%%%%%%%%%%%%%%%%%%%%%%%%%%%%%%%%%%%%%%%%%%%%%%%%%%%%%%%%%%%%%

\printindex
\end{document}